\documentclass[11pt,a4paper,fleqn]{article}

\usepackage[ansinew]{inputenc}
\usepackage[mathscr]{eucal}
\usepackage{amsmath,amssymb,amsthm}
\usepackage{graphicx}
\usepackage{caption}
\usepackage{subcaption}
\usepackage{hyperref}

\allowdisplaybreaks
\hypersetup{colorlinks, linkcolor=blue, citecolor=blue, urlcolor=blue}

\makeatletter
\long\def\@makecaption#1#2{%
  \vskip\abovecaptionskip\footnotesize
  \sbox\@tempboxa{#1. #2}%
  \ifdim \wd\@tempboxa >\hsize
    #1. #2\par
  \else
    \global \@minipagefalse
    \hb@xt@\hsize{\hfil\box\@tempboxa\hfil}%
  \fi
  \vskip\belowcaptionskip}
\makeatother

\newcommand{\todo}[1][\null]{\ensuremath{\clubsuit}}

\newcommand{\noprint}[1]{}
\newcommand{\checked}[1][\null]{\ensuremath{\boldsymbol{\surd}}}

{\theoremstyle{definition}

\newtheorem{remark}{Remark}
\newtheorem*{remark*}{Remark}
}

\newcommand{\p}{\partial}

\newcommand{\DD}{\mathrm{D}}
\newcommand{\vv}{\mathbf{v}}
\newcommand{\ve}{\varepsilon}
\newcommand{\nn}{\nabla}

\newcommand{\ZZ}{\mathcal{Z}}

\newcommand{\XX}{\mathcal{X}}
\newcommand{\DDD}{\mathcal{D}}

\begin{document}

\par\noindent {\LARGE\bf
Invariant parameterization of geostrophic eddies\\ in the ocean
\par}

\vspace{6mm}\par\noindent{\bf
Alexander Bihlo$^{\dag}$, Elsa Dos Santos Cardoso-Bihlo$^{\dag}$ and Roman O.\ Popovych$^{\ddag}$
}

\vspace{6mm}\par\noindent{\it
$^\dag$\,Department of Mathematics and Statistics, Memorial University of Newfoundland,\\
$\phantom{^\dag}$\,St.\ John's (NL) A1C 5S7, Canada
}

\vspace{2mm}\par\noindent{\it
$^\ddag$\,Fakult\"at f\"ur Mathematik, Universit\"at Wien, Oskar-Morgenstern-Platz 1, A-1090 Wien, Austria%
\\
$\phantom{^\ddag}$Institute of Mathematics of NAS of Ukraine, 3 Tereshchenkivska Str., 01601 Kyiv, Ukraine
}

\vspace{6mm}\par\noindent
E-mails:
abihlo@mun.ca,
ecardosobihlo@mun.ca,
rop@imath.kiev.ua

\vspace{9mm}\par\noindent\hspace*{8mm}\parbox{140mm}{\small\looseness=-1
The framework of invariant parameterization is extended to higher-order closure schemes.
We also define, for the first time, generalized invariant parameterization schemes,
where symmetries of the corresponding original model are preserved as equivalence transformations of related classes of closed system of differential equations.
As a particular problem, we consider invariant parameterization schemes for geostrophic eddies in a barotropic ocean. Here the initial model is the barotropic vorticity equation, which is equivalent to the system of incompressible inviscid two-dimensional Euler equations on a midlatitude beta-plane. The maximal Lie invariance algebra of this model is infinite-dimensional, and we intend to preserve it in the course of invariant parameterization, at least partially. The parameterizations proposed for the eddy vorticity flux and the energy flux are of order one and a half since we explicitly consider the equation for the turbulent kinetic energy in the closure models. These parameterizations are therefore the first examples of invariant higher-order closure schemes. Numerical experiments are carried out to assess the performance of these invariant schemes in studies of freely decaying two-dimensional turbulence, and it is verified that the invariant parameterization schemes give, on average, better results than the standard non-invariant closure models do.

}\par\vspace{5mm}

\section{Introduction}

Averaging a nonlinear model of fluid motion leads to the \emph{closure problem}. There are terms arising in the averaged model that cannot be specified using just the information contained in the mean prognostic variables. In order to close the differential equations for these dependent variables, a subgrid-scale closure model has to be employed, i.e.\ it is necessary to specify a relation between the unresolved subgrid-scale values and the resolved mean fields. Such a subgrid-scale closure model parameterizes the effects of the unresolved subgrid-scale values, e.g., the unresolved geostrophic ocean eddies, in the averaged model.

As reconstructing these effects only from the information contained in the mean fields themselves can never be perfectly accurate, there is no single canonical closure model. Nevertheless, there are a few commonly accepted general parameterization strategies, which are formulated, e.g., in~\cite{stul88Ay}. Such strategies include maintaining of the correct dimensionality of the parameterized terms, the preservation of tensorial properties or invariance of the closure model under the change of the coordinate systems, just to name a few.

Unfortunately, these parameterization strategies are seldom sufficient to obtain a unique closure model for the subgrid-scale processes. This is why in~\cite{popo10Cy} a new strategy was proposed to restrict the multitude of parameterization models using geometric properties of differential equations, which is the framework of \emph{invariant parameterization}. 
The problem of invariant parameterization is a quite natural one. Physical parameterization schemes are often designed to be isotropic or homogeneous in which case they are naturally rotationally or shift invariant. This mimics various physical processes that may be taking place in isotropic or homogeneous media such as constant-density or layered fluids. 

The motivation to tackle the invariant-parameterization problem emerged from the work by Speziale, who emphasized the importance of Galilean invariance in subgrid-scale closure models for the filtered Navier--Stokes equations~\cite{spez85Ay}. Later, Oberlack extended the work of Speziale by establishing conditions that enable the construction of general subgrid-scale closure models for the filtered Navier--Stokes equations that respect all the Lie symmetries admitted by the unfiltered Navier--Stokes equations~\cite{ober97Ay}. In practice, this leads to conditions both on the filter kernel to be used for the filtering and on the form of the subgrid-scale closure that can be taken.
While physically different, we use the notions of averaging and filtering interchangeably, since the theory below can be developed for both operations at the same time.

The systematic extension of this idea to the construction of general \emph{invariant parameterization schemes} for subgrid-scale terms of \emph{any} averaged model was given in~\cite{popo10Cy}, see also~\cite{bihl11Fy}. The main contribution here was to identify the problem of invariant parameterization as a group classification problem. Group classification aims to describe symmetries for system of differential equations from certain classes, i.e.\ for systems that include  constant or functional parameters. By assuming a functional relation between the unknown subgrid-scale terms to be parameterized and the known grid-scale quantities to be used for the parameterization scheme and plugging this ansatz into the averaged or filtered governing equations, these equations then readily constitute a class of differential equations. In other words, the invariant-parameterization problem is brought into the form of a group classification problem, that can subsequently be tackled using standard methods from the field of group analysis of differential equations, such as the algebraic method of group classification \cite{bihl11Dy,bihl17c,opan17a,popo06Ay,popo10Cy,popo10Ay}.

A similar strategy was pursued in \cite{bihl12Dy,bihl11Fy,popo17a} to find general parameterization schemes that preserve conservation laws of initial systems of differential equations in the corresponding averaged models and are hence called \emph{conservative parameterization schemes}.

From the practical point of view, the only models so far investigated using the methods of invariant parameterization employ first-order closure schemes. That is, for the parameterization of the unknown subgrid-scale quantities, only the grid-scale variables and their derivatives have been used. In view of state-of-the-art applications, first-order (local) closure models are certainly inadequate, as there is no way to include information on, e.g., variances of turbulent perturbation quantities in such models. The reason for this is that these variances are at once replaced by the grid-scale quantities, which is more than often insufficient from a physical point of view.

The present paper seeks to overcome the above problem by considering so-called one-and-a-half order closure schemes~\cite{stul88Ay}. The idea of them is to close the averaged system of differential equations by also taking into account a part of the variance of the turbulent perturbation quantities; here, specifically, we will consider closures that explicitly contain equations for the turbulent kinetic energy. This energy is a measure for the intensity of turbulence and as such is an important quantity in turbulence research and in the problem of eddy parameterization in both meteorology and oceanography. From the point of view of invariant parameterization, retaining the turbulent kinetic energy is a proof-of-the-concept that preserving symmetries in closure schemes is not restricted to the lowest-order parameterization models and can successfully be extended to higher-order models as well.

The second novel feature of the paper is the generalized notion of invariant parameterization. 
Usually, a closure scheme involves parameters, i.e., the closure of an averaged system of differential equations 
leads not to a single closed system but to a class of such systems, 
where the tuple of arbitrary elements of the class is constituted by the closure parameters.
The parameterization process can be interpreted as \emph{symmetry-preserving in the generalized sense}
if the obtained class admits simultaneous prolongations of selected Lie symmetries of the original system to subgrid-scale quantities and to the closure parameters
as its equivalence transformations.

\looseness=-1
The starting model is the system of two-dimensional incompressible Euler equations on a mid-latitude beta-plane, which we represent in the form of the single equation for the stream function, the so-called barotropic vorticity equation on the beta-plane. This model is of superior importance in atmosphere--ocean dynamics as it takes into account the quasi-horizontality of the fluid system as well as the driving rotational effects. Despite its relative simplicity, it is one of the standard models in ocean modeling and a prototype for testing new parameterization methodologies \cite{bihl11Fy,cumm92a,davi17a,eden08Ay,grea00a,mars10Ay,wang94a}.
We study the problem of parameterizing eddies in a barotropic ocean model in a manner so as to preserve, in a certain sense, Lie symmetries of the original model. 
Carried out numerical experiments show that the performance of the invariant parameterization schemes is, on average, better than that of the standard non-invariant closure models. 

The further organization of this paper is the following. 
We begin by presenting and extending the theory of invariant parameterization in Section~\ref{sec:MethodInvariantParameterization}. 
In Section~\ref{sec:barotropicOcean} we describe the barotropic ocean model that was studied in~\cite{mars10Ay} and that we are going to use. In Section~\ref{sec:symmetriesBarotropicOcean} we recall the maximal Lie symmetry group~$G$ of the barotropic vorticity equation on the beta-plane
and show how to prolong transformations from~$G$ to the turbulent kinetic energy. 
Lie symmetries of the incompressible inviscid two-dimensional Euler equations on a midlatitude beta-plane are computed for the first time and compared with those of the barotropic vorticity equation on the beta-plane. As a result, a genuinely potential symmetry of the former equations is found. 
Differential invariants of the group~$G$ are constructed in Section~\ref{sec:invariantParameterizationBarotropicOcean}, which forms the basis for finding invariant parameterization schemes for geostrophic eddies in the barotropic ocean model. Some numerical results on the new invariant parameterization schemes and on the standard non-invariant closure models are presented and compared in Section~\ref{sec:numericalResultsBarotropicOcean}. In the final Section~\ref{sec:conclusionBarotropicOcean} we sum up the results of the paper and give some thoughts for future research directions on the problem under consideration.

\section{The method of invariant parameterization}\label{sec:MethodInvariantParameterization}

The first systematic theory on invariant parameterization was laid out in~\cite{popo10Cy}. There it was argued that the problem of invariant parameterization is essentially equivalent to a (complicated) problem of group classification. Additional details can be found in~\cite{bihl11Fy,bihl12Dy}. We extend this idea here to higher-order parameterization schemes. 

Suppose we are given a system of differential equations of the form
\begin{equation}\label{eq:GenericSystemOfDE}
\Delta^l(x,u_{(n)})=0,\quad l=1,\dots,m.
\end{equation}
Here and in the following we denote by $x=(x^1,\dots,x^p)$ the independent variables and by $u_{(n)}$ the dependent variables $u=(u^1,\dots,u^q)$ jointly with all derivatives of $u$ with respect to $x$ up to order~$n$, 
and $\Delta^l$ are sufficiently smooth functions of~$(x,u_{(n)})$.
We use the subscript notation to denote differentiation with respect to the associated variables.

When preparing system~\eqref{eq:GenericSystemOfDE} for a numerical implementation, one first chooses a suitably fine grid on which the numerical solution will live. Effectively, this means that the unknown functions~$u$ are decomposed into $u=\bar u+u'$, where $\bar u$ is the resolved part of $u$ (the grid-scale part) living on the discretization mesh, and $u'$ is the unresolved part of $u$ (the subgrid-scale part). Thus, practically speaking, $\bar u$ can be computed using a numerical model for system~\eqref{eq:GenericSystemOfDE}, whereas $u'$ cannot be resolved by the model.

An alternative interpretation is that we regard $\bar u$ as the function obtained by averaging (or filtering) $u$ over each grid cell. Various averaging or filtering operations can be used, but in the following we restrict ourselves to classical \textit{Reynolds averaging}. The Reynolds averaging satisfies the Reynolds property $\overline{\bar ab}=\bar a\bar b$, which implies the \textit{Reynolds averaging rule} $\overline{ab}=\bar a\bar b+\overline{a'b'}$. Here the bar is to be interpreted as averaging in both space and time. We should like to stress though that the precise form of averaging (filtering) or discretization methodology chosen to obtain a system of equations for $\bar u$ is not essential from the theoretical point of what will follow.

To implement system~\eqref{eq:GenericSystemOfDE} numerically we thus have to convert it into a system for $\bar u$. Assuming system~\eqref{eq:GenericSystemOfDE} to be nonlinear, Reynolds-averaging it yields a system of the form
\begin{equation}\label{eq:AveragedSystemUnclosed}
\bar\Delta^{1l}\big(x,\bar u_{(n)}^{},w^1_{(n')}\big)=0,\quad l=1,\dots,m,
\end{equation}
where $\bar\Delta$'s are sufficiently smooth functions of their arguments. The precise form of these functions and the maximal order~$n'$ of involved derivatives of~$w^1$ depends on the equations in the original system~\eqref{eq:GenericSystemOfDE} as well as on the averaging rule invoked. The tuple $w^1=(w^{11},\dots,w^{1j})$ collects all subgrid-scale terms originating from applying the averaging rule to nonlinear combinations of terms in the original system~\eqref{eq:GenericSystemOfDE}. These terms are those that need to be expressed via the resolved quantities in order to close the system~\eqref{eq:AveragedSystemUnclosed}.

The most straightforward way of closing~\eqref{eq:AveragedSystemUnclosed} is to make an ansatz 
\[
w^{1i}=h^i(x,\bar u_{(r)}),\quad i=1,\dots,j, 
\]
where $h=(h^1,\dots,h^j)$ are parameterization functions, which should be specified, $r\in\mathbb N\cup\{0\}$. The associated parameterization scheme is \textit{local} and of \textit{first order}~\cite{stul88Ay}. System~\eqref{eq:AveragedSystemUnclosed} is closed to
\begin{equation}\label{eq:AveragedSystemClosed}
\bar\Delta^{1l}\big(x,\bar u_{(n)},h_{[n']}(x,\bar u_{(r)})\big)=0,\quad l=1,\dots,m,
\end{equation}
which is a class of differential equations, where $h$ plays the role of the tuple of the arbitrary elements. 
Here and in what follows we denote by $h_{[n']}$ the collection of~$h$ and all total derivatives of $h$ with respect to $x$ up to order~$n'$.
Thus, the orders of equations in systems of the form~\eqref{eq:AveragedSystemClosed} is not greater than $\bar n:=\max(n,n'+r)$.
Symmetry-preserving parameterization schemes are found upon solving a \textit{group classification problem} for the class of systems of the form~\eqref{eq:AveragedSystemClosed}. There are two main paradigms for solving a group classification problem, using \textit{direct group classification} or \textit{inverse group classification}.

Direct group classification proceeds by investigating for which specific values of the parameterization functions~$h$, systems from the class~\eqref{eq:AveragedSystemClosed} admit more symmetries than in the case of generic~$h$. To facilitate an efficient classification, it is necessary to compute the so-called \textit{equivalence group}~$G^\sim$ of the class~\eqref{eq:AveragedSystemClosed} first. Elements of the group~$G^\sim$, which are called equivalence transformations of the class~\eqref{eq:AveragedSystemClosed}, are point transformations in the extended space with coordinates $(x,u_{(\bar n)},h)$ that map systems from this class to systems of the same class. The classification of values of the parameter functions~$h$ such that the associated systems admit more symmetries than the generic case is then done up to $G^\sim$-equivalence. Several methods have been proposed to efficiently solve the group classification problem, among which the \textit{algebraic method} is the most powerful for the classes of differential equations that typically arise within the problem of invariant parameterization. See~\cite{bihl11Dy,bihl17c,popo10Cy,opan17a,popo10Ay} for the use of the algebraic method to solving group classification problems.

Inverse group classification of the class~\eqref{eq:AveragedSystemClosed} relies on fixing a specific local Lie group~$G$ that acts in the space with the coordinates $(x,u)$ 
and further looking for systems from the class~\eqref{eq:AveragedSystemClosed} that are invariant with respect to the group~$G$. 
The crucial step is to construct $\bar n$th order \textit{differential invariants} of~$G$, i.e., functions of~$(x,u_{(\bar n)})$ that are invariant with respect to the prolongation of~$G$ to derivatives of~$u$ up to order~$\bar n$. This prolongation is computed by the chain rule. 
Differential invariants are combined together to yield $G$-invariant systems of at most $\bar n$th order differential equations. 
In the context of the invariant-parameterization problem, inverse group classification is the approach originally used in~\cite{ober97Ay}, where it was determined that any subgrid-scale closure scheme for the Navier--Stokes equations should admit the same symmetries as the original Navier--Stokes equations do. In practice, invariant parameterization schemes are most conveniently computed within the framework of inverse group classification using the method of equivariant moving frames~\cite{fels98Ay,fels99Ay,cheh08Ay,olve08Ay}, which allows one to readily obtain differential invariants using an \textit{invariantization map}. In~\cite{bihl11Fy}, it was proposed that invariant parameterization schemes can be constructed by applying a suitable invariantization map directly to a given closed, non-invariant system of the form~\eqref{eq:AveragedSystemClosed}; cf.\ Section~\ref{sec:invariantParameterizationBarotropicOcean}.

It can also make sense to relax the assumption that the closed model~\eqref{eq:AveragedSystemClosed} should preserve all the symmetries of the original model~\eqref{eq:GenericSystemOfDE}. In particular, the closed model~\eqref{eq:AveragedSystemClosed} only captures the resolved part of the dynamics of the original model~\eqref{eq:GenericSystemOfDE} and thus requiring invariance of this model under exactly the same symmetries as admitted by the parent model may be overly restrictive or even unphysical. The presence of particular boundary conditions imposed on the closed model~\eqref{eq:AveragedSystemClosed} may also yield natural restrictions on which symmetries should be preserved by this model, see e.g.~\cite{bihl11By,bihl12By,blum89Ay}. Lastly, not all symmetries of the original model~\eqref{eq:GenericSystemOfDE} may be physically relevant to begin with. 
An example for the last point is provided by the system of (1+1)-dimensional shallow-water equations whose maximal Lie invariance algebra is infinite-dimensional, whereas the maximal Lie invariance algebra of its (1+2)-dimensional counterpart is finite-dimensional only. The infinite dimension of the former algebra is an artefact since the system of (1+1)-dimensional shallow-water equations, being a system of two first-order quasilinear equations in two independent variables, can be linearized using a rank-two hodograph transformation.
Thus, in constructing invariant parameterization schemes for the (1+1)-dimensional shallow-water equations, one should restrict oneself to preserving at most those Lie symmetries that are counterparts of symmetries of the (1+2)-dimensional shallow-water equations, see~\cite{bihl12By} for further discussions.

In the sequel we will use the inverse classification approach, building upon the results obtained in~\cite{bihl11Fy} for the barotropic vorticity equation on the beta-plane. The main novel feature is the use of a higher-order parameterization scheme to close the eddy-vorticity flux term that arises in the course of Reynolds-averaging the vorticity equation.

In general, higher-order parameterization schemes aim to improve upon the overly simplistic assumption that the unresolved subgrid-scale fields $w^1$ can be directly expressed in terms of the resolved grid-scale quantities $\bar u_{(r)}$. Instead, one proceeds by deriving equations for the subgrid-scale fields $w^1$ themselves. The resulting equations will again be unclosed, depending on higher-order subgrid-scale fields $w^2$ (triple correlation quantities, such as $\overline{u'u'_xu'_{xx}}$). One can repeat this construction and thus derive a system of the form
\begin{gather}\label{eq:AveragedSystemUnclosedSOrder}
\bar\Delta^{\varsigma l_\varsigma}\big(x,\bar u_{(n)},w^1_{(n'_\varsigma)},\dots,w^\varsigma_{(n'_\varsigma)}\big)=0,\quad l_\varsigma=1,\dots,m_\varsigma,\quad \varsigma=1,\dots,s,
\end{gather}
where $w^\varsigma=(w^{\varsigma1},\dots,w^{\varsigma j_\varsigma})$, $\varsigma=1,\dots,s$,
$j_\varsigma,m_\varsigma,n'_\varsigma\in\mathbb N$, $j_1:=j$, $m_1:=m$, $n'_1:=n'$. 
Note that this system is still not closed, as there are no prognostic equations for $w^s=(w^{s1},\dots,w^{sj_s})$. 
A~\textit{local $s$th order parameterization scheme} is obtained by setting an ansatz
\begin{gather}\label{eq:SOrderClosure}
w^{si}=h^i\big(x,\bar u_{(r)}^{},w^1_{(r)},\dots,w^{s-1}_{(r)}\big),\quad i=1,\dots,j_s,
\end{gather} 
for some $r\in\mathbb N\cup\{0\}$ and some tuple of sufficiently smooth functions $h=(h^1,\dots,h^{j_s})$ of their arguments,
which will close system~\eqref{eq:AveragedSystemUnclosedSOrder} and are called \textit{parameterization functions}, yielding
\begin{equation}\label{eq:AveragedSystemClosedSOrder}
\begin{split}
&\bar\Delta^{\varsigma l_\varsigma}\big(x,\bar u_{(n)}^{},w^1_{(n'_\varsigma)},\dots,w^\varsigma_{(n'_\varsigma)}\big)=0,\quad l_\varsigma=1,\dots,m_\varsigma,\quad \varsigma=1,\dots,s-1,\\
&\bar\Delta^{sl_s}\big(x,\bar u_{(n)}^{},w^1_{(n'_s)},\dots,w^{s-1}_{(n'_s)},h_{[n'_s]}^{}\big(x,\bar u_{(r)}^{},w^1_{(r)},\dots,w^{s-1}_{(r)})\big)=0.
\end{split}
\end{equation}
We call a local $s$th order parameterization scheme~\eqref{eq:SOrderClosure} with a specific~$h$ \textit{invariant} if the system~\eqref{eq:AveragedSystemClosedSOrder} with this~$h$ is invariant under a prolongation of the maximal Lie invariance group (or a suitable subgroup) of system~\eqref{eq:GenericSystemOfDE} to the subgrid-scale fields $w^1$, \dots, $w^{s-1}$ or, more generally, under a reasonable point transformation group acting in the space with the coordinates $(x,\bar u,w^1,\dots,w^{s-1})$. The prolongation of the group action originally defined in the space with coordinates $(x,u)$ to the unresolved terms $(w^1,\dots,w^{s-1})$ will be illustrated conceptually for the barotropic vorticity equation in the following section. 
From the point of view of symmetry analysis of differential equations, 
the construction of invariant local $s$th order parameterization schemes reduces to solving 
a group classification problem for systems for the class~\eqref{eq:AveragedSystemClosedSOrder}, 
where $h$ plays the role of the tuple of the arbitrary elements, 
and the orders of differential equations in systems is not greater than $\bar n:=\max\{n,n'_\varsigma,\varsigma=1,\dots,s-1,n'_s+r\}$.

We can also interpret the notion of invariant parameterization in a more general sense.
Suppose that in the class~\eqref{eq:AveragedSystemClosedSOrder}, one singles out a sufficiently narrow subclass~$\mathscr P$
that can be assumed, after a re-parameterization, to be parameterized by constants (or functions of a few arguments), 
which are closure's parameters. 
These parameters are then considered as the arbitrary elements of the subclass~$\mathscr P$.
We call a local $s$th order parameterization scheme~\eqref{eq:SOrderClosure}, 
where the tuple~$h$ runs though the set associated with the subclass~$\mathscr P$, \textit{invariant in the generalized sense} 
if the equivalence group of~$\mathscr P$ contains a simultaneous prolongation of the maximal Lie invariance group (or a suitable subgroup) of system~\eqref{eq:GenericSystemOfDE} to the subgrid-scale fields $w^1$, \dots, $w^{s-1}$ and to the arbitrary elements of~$\mathscr P$
or, more generally, is a reasonable point transformation group acting in the joint space of the variables $(x,\bar u,w^1,\dots,w^{s-1})$ and the arbitrary elements of~$\mathscr P$.
Of course, some of relevant equivalence transformations can be associated, via the projection to the space with the coordinates $(x,\bar u,w^1,\dots,w^{s-1})$,
with common symmetries of systems from the subclass~$\mathscr P$.

\section{The model of a barotropic ocean}\label{sec:barotropicOcean}

We start with the model of the barotropic ocean as given through the barotropic vorticity equation on the beta-plane:
\begin{equation}\label{eq:barotropicVorticityEquation}
 \zeta_t + \psi_x\zeta_y-\psi_y\zeta_x+\beta\psi_x = 0,\qquad \textup{or}\qquad \eta_t + \psi_x\eta_y-\psi_y\eta_x = 0,
\end{equation}
where $\eta = \zeta+{\rm f}_0+\beta y= \psi_{xx}+\psi_{yy}+{\rm f}_0+\beta y$ is the absolute vorticity, which is the sum of the relative vorticity $\zeta=\psi_{xx}+\psi_{yy}$ and the affine approximation ${\rm f}_0+\beta y$ of the Coriolis parameter, which is called the beta-plane approximation and incorporates the linear change of the Coriolis parameter in the North--South direction. The constants~${\rm f}_0$ and~$\beta$ are the value of the Coriolis parameter on a fixed geographical latitude and the rate of the above linear change, respectively, and $\beta\ne0$.

The vorticity equation on the beta-plane~\eqref{eq:barotropicVorticityEquation} can be derived from the system of incompressible two-dimensional Euler equations in a rotating reference frame in the absence of external forces. This system consists of the momentum equations and the incompressibility condition,
\begin{subequations}\label{eq:EulerEqs}
\begin{gather}\label{eq:MomentumEqs}
\vv_t+(v_x-u_y+{\rm f}_0+\beta y)\vv^\bot + \nn B=0,
\\\label{eq:IncompressibilityCondition}
\nn\cdot \vv=0,
\end{gather}
\end{subequations}
where $\vv=(u,v)^{\mathsf T}$ is the two-dimensional velocity field, $\vv^\bot=(-v,u)^{\mathsf T}$, and $B=p/\rho+\vv^2/2$ is the Bernoulli function, given as the sum of the mass-specific potential energy and the mass-specific kinetic energy, $p$ is the pressure and $\rho$ is the (constant) density. The stream function and the velocity field are related through $\vv = (\nn\psi)^\bot$, i.e., $u=-\psi_y$ and $v=\psi_x$.

Adding the product of the incompressibility condition~\eqref{eq:IncompressibilityCondition} by~$B$ to the scalar product of the momentum equations~\eqref{eq:MomentumEqs} by $\vv$,
one obtains the kinetic energy equation in the form of the following conservation law
\begin{equation}\label{eq:kineticEnergyEquations}
 E_t+ \nn\cdot (B\vv)=0,
\end{equation}
where $E=\vv^2/2=(\nn\psi)^2/2$ is the kinetic energy. We will need to use this equation in the construction of invariant parameterization schemes since we will take into account not only the vorticity and its derivatives but also the turbulent kinetic energy. 
This will be an example of a \textit{one-and-half order parameterization scheme} in view of the fact that explicit equations are merely provided for a subset of all second-order correlation quantities, see e.g.~\cite{stul88Ay}.

The Reynolds-averaged equations for the barotropic ocean are derived by using the decomposition $\psi(t,x,y)=\bar \psi(t,x,y) +\psi'(t,x,y)$, where $\bar \psi$ is the resolved part of the stream function and $\psi'$ is the perturbation of the mean. Then, Reynolds-averaging the vorticity equation~\eqref{eq:barotropicVorticityEquation} leads to
\begin{align}\label{eq:averagedVorticityEquation}
  &\bar\eta_t + \bar\psi_x\bar\eta_y-\bar\psi_y\bar\eta_x = \nn\cdot(\overline{\vv'\eta'}),
\end{align}
where $\bar\eta=\bar\zeta+{\rm f}_0+\beta y=\bar\psi_{xx}+\bar\psi_{yy}+{\rm f}_0+\beta y$, see also~\cite{bihl11Fy,popo10Cy}. In deriving equation~\eqref{eq:averagedVorticityEquation} we used the Reynolds-averaging property that $\overline{ab}=\bar a\bar b+\overline{a'b'}$ for any product of two dependent variables.

The parameterization problem now consists of finding an expression for the right-hand side of~\eqref{eq:averagedVorticityEquation}, which physically constitutes the divergence of the eddy vorticity flux. In~\cite{bihl11Fy,popo10Cy}, we solved this problem by assuming a functional relation between the eddy vorticity flux and the derivatives of the averaged stream function $\bar \psi$. Here we will assume that this functional relation also involves the \emph{turbulent kinetic energy}, $k=\overline{\vv'^2}/2=\overline{(\nn\psi')^2}/2$.

To be able to use the turbulent kinetic energy and its derivatives as resolved quantities, it is necessary to formulate an explicit evolution equation for the turbulent kinetic energy and attach it to~\eqref{eq:barotropicVorticityEquation}. This equation is obtained by first considering the momentum equation for the perturbation velocity $\vv'$, which is
\[
 \vv'_t + (\eta\vv'+\eta'\vv+\eta'\vv')^\bot+\nn B'=0,
\]
where $\eta'=\zeta'$ since the Coriolis parameter is a part of the mean absolute vorticity. Then, taking the scalar product of this equation with $\vv'$ and averaging the result, we obtain the equation for the turbulent kinetic energy
\begin{equation}\label{eq:turbulentKineticEnergyEquation}
 k_t -\vv\cdot(\overline{\eta'\vv'})^\bot+\nn\cdot\overline{B'\vv'}=0,
\end{equation}
see also~\cite{mars10Ay}.

In order to close the system of two equations~\eqref{eq:averagedVorticityEquation} and~\eqref{eq:turbulentKineticEnergyEquation} we need to parameterize three terms,
$\nn\cdot(\overline{\vv'\eta'})$, $\vv\cdot(\overline{\eta'\vv'})^\bot$ and $\nn\cdot\overline{B'\vv'}$. We aim to do this by finding appropriate functional relations between these terms and $\bar\psi$, $\bar\eta$, $k$ and their derivatives. The explicit occurrence of the turbulent kinetic energy $k$ is what will make the respective parameterization schemes of order one-and-a-half. Moreover, we require that the closed version of the system~\eqref{eq:averagedVorticityEquation}--\eqref{eq:turbulentKineticEnergyEquation} will possess the counterparts of selected symmetries admitted by the original barotropic vorticity equation~\eqref{eq:barotropicVorticityEquation}. This is in accordance with the general theory of invariant parameterization schemes as outlined above and in~\cite{popo10Cy}.

\begin{remark}
It is important to stress that in order to find invariant closure schemes for the system of averaged equations~\eqref{eq:averagedVorticityEquation} and~\eqref{eq:turbulentKineticEnergyEquation} it is sufficient to only take into account the symmetries of the original vorticity equation~\eqref{eq:barotropicVorticityEquation}. The reason for this is that the energy equation is a \emph{differential consequence} (more precisely, a conservation law) of the Euler equations~\eqref{eq:EulerEqs} or, equivalently, of the vorticity equation~\eqref{eq:barotropicVorticityEquation}.
Hence, the only really independent piece of information in this problem is given by the system~\eqref{eq:EulerEqs} or, equivalently, by the vorticity equation~\eqref{eq:barotropicVorticityEquation}.
\end{remark}

\section{Symmetries of the barotropic ocean model}\label{sec:symmetriesBarotropicOcean}

The Lie symmetries of the barotropic vorticity equation on the beta-plane~\eqref{eq:barotropicVorticityEquation} are well studied, see e.g.~\cite{katk65Ay,katk66Ay} for a first description and~\cite{bihl09Ay} for some more recent work. The maximal Lie invariance algebra~$\mathfrak g$ of the equation~\eqref{eq:barotropicVorticityEquation} is infinite dimensional and spanned by the vector fields
\begin{align}\label{eq:symmetryGroupVorticityEquation}
\begin{split}
&\DDD=t\p_t-x\p_x-y\p_y-3\psi\p_\psi,\quad 
\p_t,\quad \p_y,\\
&\XX(f)=f(t)\p_x-f_t(t)y\p_\psi,\quad 
\ZZ(g)=g(t)\p_\psi.
\end{split}
\end{align}
Here and in what follows the parameters~$f$ and~$g$ run through the set of real-valued smooth time-dependent functions. As usual, the shorthand notation like $\p_t=\partial/\partial t$ is used to abbreviate partial derivative operators. Therefore, elements of the maximal Lie symmetry group~$G$ of the equation~\eqref{eq:barotropicVorticityEquation} are point transformations of the form
\begin{equation}\label{eq:GeneralSymmetryTransformationVorticityEquation}
 (\tilde t,\tilde x,\tilde y,\tilde \psi) = (e^{\ve_3}(t+\ve_1),e^{-\ve_3}(x+f(t)),e^{-\ve_3}(y+\ve_2), e^{-3\ve_3}(\psi-f_t(t)y+g(t)) )
\end{equation}
where $\ve_1$, $\ve_2$ and $\ve_3$ run through the set of real numbers.

Since the barotropic vorticity equation is derived from the two-dimensional incompressible Euler equations~\eqref{eq:EulerEqs}, it is natural to also present the Lie symmetries of the latter model. To the best of our knowledge, this has not been done in the literature before.
The maximal Lie invariance algebra of~\eqref{eq:EulerEqs} is spanned by the vector fields
\begin{align}\label{eq:symmetryGroupEulerEquation}
\begin{split}
&\hat\DDD=t\p_t-x\p_x-\left(y+\frac{{\rm f}_0}\beta\right)\p_y-2u\p_u-2v\p_v-4p\p_p,\quad 
\p_t,\\
&\hat\XX(f)=f(t)\p_x+f_t(t)\p_u-\rho\left(f_{tt}(t)x+\frac\beta2f_t(t)y^2+{\rm f}_0f_t(t)y\right)\p_p,\quad 
\hat\ZZ(g)=g(t)\p_p.
\end{split}
\end{align}
It is instructive to compare the algebras~\eqref{eq:symmetryGroupVorticityEquation} and~\eqref{eq:symmetryGroupEulerEquation}. 
The vector fields~$\DDD-{\rm f}_0\beta^{-1}\p_y$, $\p_t$ and~$\XX(f)$ in~$\mathfrak g$ are induced by the vector fields~$\hat\DDD$, $\p_t$ and~$\hat\XX(f)$ in \eqref{eq:symmetryGroupEulerEquation}, respectively. The appearance of the vector fields~$\ZZ(g)$ among the spanning elements~\eqref{eq:symmetryGroupVorticityEquation} of the algebra~$\mathfrak g$ is a direct consequence of introducing the stream function~$\psi$ as a potential in place of the velocity vector~$\vv$ using the expressions for $u$ and $v$ in terms of~$\psi$, which solve the incompressibility condition~\eqref{eq:IncompressibilityCondition}. 
Eliminating the pressure term in the Euler equations by cross differentiation (for reformulating the Euler equations merely in terms of the stream function) 
leads to that the vector fields $\hat\ZZ(g)$ from~\eqref{eq:symmetryGroupEulerEquation} have no counterparts among the vector fields~\eqref{eq:symmetryGroupVorticityEquation}. 
The most interesting observation is the absence of a counterpart of the single vector field~$\p_y$ in~\eqref{eq:symmetryGroupEulerEquation}. 
The intermediate system of equations obtained by introducing the stream function~$\psi$ in~\eqref{eq:EulerEqs}, which is called a potential system for~\eqref{eq:EulerEqs},
admits $\p_y+\rho\beta\psi\p_p$ as a Lie symmetry vector field. 
Thus, this vector field is a so-called potential symmetry for the original system~\eqref{eq:EulerEqs}.
After eliminating the pressure~$p$ from the intermediate system, which is done in the course of deriving the barotropic vorticity equation, this vector field reduces to~$\p_y$.
The fact that shifts of an independent variable induce potential symmetries is rather unusual. 
Shifts with respect to the variable~$y$ are quite natural symmetry transformations for models with zonally homogeneous Coriolis force. 
Converting these important Lie symmetries of~\eqref{eq:barotropicVorticityEquation} to potential symmetries of~\eqref{eq:EulerEqs}  
can be seen as a symmetry argument for working with the vorticity equation~\eqref{eq:barotropicVorticityEquation} instead of the Euler equations~\eqref{eq:EulerEqs} 

The characteristic of one-and-a-half order closure schemes is the explicit presence of a quantity like $k$, the turbulent kinetic energy. 
As was argued in the previous section, the counterparts of selected symmetries of the vorticity equation in terms of the mean stream function and turbulent kinetic energy should be preserved in an invariantly closed version of the system~\eqref{eq:averagedVorticityEquation}--\eqref{eq:turbulentKineticEnergyEquation}. 
In order to include the turbulent kinetic energy in this problem, i.e., to make the differential invariants to be derived explicitly dependent on~$k$, we need to prolong the vector fields given in~\eqref{eq:symmetryGroupVorticityEquation} to~$k$. This is readily accomplished by inferring the transformation rules for $k$, from its definition, $k=\overline{(\nn\psi')^2}/2$. In other words, the transformation behavior for $k$ follows from that of~$\psi_x'$ and~$\psi_y'$, which is readily obtained from the most general symmetry transformation~\eqref{eq:GeneralSymmetryTransformationVorticityEquation}.

Specifically, consider the splitting
$
 \psi_x=\bar\psi_x+\psi_x'
$
and determine the transformation behavior of the right-hand side. We have that
\[
 \widetilde{\psi_x}=\tilde\psi_{\tilde x}=e^{-2\ve_3}\psi_x =e^{-2\ve_3}(\bar\psi_x+\psi_x')=\widetilde{\bar\psi_x}+\widetilde{\psi_x'}
\]
from which it is readily inferred that
\[
 \widetilde{\bar\psi_x} = e^{-2\ve_3}\bar \psi_x,\quad \widetilde{\psi_x'}=e^{-2\ve_3}\psi_x'.
\]
Similarly, for $\psi_y$ we have
\[
 \widetilde{\psi_y}=e^{-2\ve_3}(\psi_y-f_t)=e^{-2\ve_3}(\bar\psi_y+\psi_y'-f_t)=\widetilde{\bar\psi_y}+\widetilde{\psi_y'}.
\]
Here we find that
\[
 \widetilde{\bar\psi_y} = e^{-2\ve_3}(\bar \psi_y-f_t),\quad \widetilde{\psi_y'}=e^{-2\ve_3}\psi_y',
\]
because we make the physically reasonable assumption that generalized Galilean transformations are acting on the value of the mean part of the flow, rather than on the value of the perturbation part.

In view of these results, it is obvious that $k$ transforms as
\[
 \tilde k = e^{-4\ve_3}k.
\]

For the sake of completeness, we have the following action of the counterpart $\bar G$ of the group~$G$ on $(t,x,y,\bar\psi,k)$:
\[
 (\tilde t,\tilde x,\tilde y,\tilde{\bar\psi},\tilde k)=(e^{\ve_3}(t+\ve_1),e^{-\ve_3}(x+f(t)),e^{-\ve_3}(y+\ve_2), e^{-3\ve_3}(\bar\psi-f_t(t)y+g(t)), e^{-4\ve_3}k).
\]

In~\cite{bihl11Fy} we have found the algebra of differential invariants for the maximal Lie invariance group of the vorticity equation in the space with coordinates $(t,x,y,\bar\psi)$. In the subsequent section we will extend this result to the action of this group on the prolonged space with coordinates $(t,x,y,\bar\psi,k)$.

\begin{remark}
In the case of higher-order parameterization schemes, see Section~\ref{sec:MethodInvariantParameterization},
prolongations of symmetry transformations to the entire tuple $(w^1,\dots,w^{s-1})$ of subgrid-scale fields with explicit prognostic equations,
could be derived in exactly the same manner. This, however, will not be pursued here.
\end{remark}

\section{Invariant parameterization for the barotropic ocean model}\label{sec:invariantParameterizationBarotropicOcean}

Invariant parameterization schemes for the vorticity equation on the beta-plane~\eqref{eq:barotropicVorticityEquation} were computed in~\cite{bihl11Fy}. In particular, therein we found a generating set of differential invariants and a complete set of independent operators of invariant differentiation for the maximal Lie symmetry group $G$ of~\eqref{eq:barotropicVorticityEquation}, i.e.\ the two ingredients needed to completely describe the algebra of differential invariants of $G$. 
We also obtain closed expressions for elements of a functional basis of differential invariants for~$G$. 
This systematic description of the algebra of differential invariants can then be used to find invariant first-order parameterization schemes for the averaged or filtered vorticity equation. Here we are going to build upon this result and generalize it to one-and-a-half order parameterization schemes.

In~\cite{bihl11Fy} we have established that the generating set of differential invariants for the group $G$ is given by the single invariant
\[
 I_1=\frac{\psi_{xx}}{\sqrt{|\psi_x|}},
\]
provided that $\psi_x\ne0$. All the other differential invariants can be constructed from~$I_1$ by combining two successive operations, (repeatedly) acting on~$I_1$ with the three independent operators of invariant differentiation
\begin{equation}\label{eq:InvariantDifferentiationOperatorsBetaPlaneEquationOcean}
 \DD^{\rm i}_t=\frac{1}{\sqrt{|\psi_x|}}(\DD_t-\psi_y\DD_x),\quad \DD^{\rm i}_x=\sqrt{|\psi_x|}\DD_x,\quad \DD^{\rm i}_y=\sqrt{|\psi_x|}\DD_y,
\end{equation}
and taking functional combinations of obtained differential invariants.
This was proved by invoking the \emph{invariantization map} given through the complete moving frame
\begin{align}\label{eq:MovingFrameBetaPlaneEquationOcean}
\begin{split}
 &\ve_1=\ln\sqrt{|\psi_x|},\quad \ve_2=-t,\quad \ve_3=-y,\quad f=-x,\\
 &f^{(i+1)}=(\DD_t-\psi_y\DD_x)^i\psi_y, \quad g^{(i)}=-(\DD_t-\psi_y\DD_x)^i\psi, \quad i=0,1,\dots,
\end{split}
\end{align}
and by explicitly evaluating the recurrence relations between the normalized and differentiated differential invariants. In~\eqref{eq:MovingFrameBetaPlaneEquationOcean} the superscripts of $f$ and $g$ stand for the number of differentiations with respect to $t$. For more details on the invariantization procedure, we refer to the literature on \emph{equivariant moving frames}~\cite{bihl11Fy,bihl17b,cheh08Ay,fels98Ay,fels99Ay,olve08Ay}.

Now considering the action of $\bar G$ on the space with coordinates $(t,x,y,\bar\psi,k)$, it is straightforward to see that we can use the invariant $I_1$ and the operators of invariant differentiation where we replace $\psi$ with $\bar \psi$. As the original $\psi$ will not appear henceforth, we omit the bars over averaged quantities for the sake of simplicity from now.

In order to have a complete basis of differential invariants of the group~$\bar G$ we need one more invariant. In particular, this invariant is obtained by invariantizing the expression for $k$, which yields the invariant
\[
 I_2 = \frac{k}{\psi_x^2}.
\]
Then $I_1$ and $I_2$ jointly with the three operators of invariant differentiation given in~\eqref{eq:InvariantDifferentiationOperatorsBetaPlaneEquationOcean} suffice to completely determine the algebra of differential invariants for the group~$\bar G$ in the extended space with coordinates $(t,x,y,\psi,k)$.

We are now in the position to turn to the problem of finding invariant parameterization schemes for the model system~\eqref{eq:averagedVorticityEquation}--\eqref{eq:turbulentKineticEnergyEquation}. 
We start this endeavor by investigating the symmetry properties of the parameterizations proposed in~\cite{mars10Ay}. These will then serve as the starting points for new closure models.

In~\cite{mars10Ay}, the following closed system stemming from the system~\eqref{eq:averagedVorticityEquation}--\eqref{eq:turbulentKineticEnergyEquation} was proposed and used:
\begin{align}\label{eq:MarshallSystem}
\begin{split}
 &\eta_t + \psi_x\eta_y - \psi_y\eta_x = \nn\cdot(\kappa\nn\eta)-A\nn^4\eta,\\
 &k_t + \psi_xk_y - \psi_yk_x = -\kappa \nn\psi\cdot\nn\eta + \nn\cdot(\nu\nn k)-rk,
\end{split}
\end{align}
where $A$ is the constant biharmonic diffusion coefficient and $\kappa$, $\nu$ and $r$ are the parameterization parameters.
The constant $\nu$ is the eddy energy diffusivity and the constant $r$ is an inverse time scale for the eddy energy decay.
For $\kappa$, which is not necessarily constant, the ansatz
\begin{equation}\label{eq:ParameterKappa}
 \kappa = \alpha L_{\rm eddy} (2k)^{1/2},
\end{equation}
where $\alpha$ is a dimensionless constant and $L_{\rm eddy}$ is a prescribed eddy mixing length scale,
was chosen in~\cite{mars10Ay}.

It is straightforward to check that as it stands, system~\eqref{eq:MarshallSystem} is invariant, for general values of involved constant parameters, 
merely with respect to a proper subgroup~$\bar G_{\rm min}$ of the group~$\bar G$, which is constituted by the transformations
\begin{gather}\label{eq:MarshallSystemGroup}
 (\tilde t,\tilde x,\tilde y,\tilde\psi,\tilde k)=(t+\ve_1,x+\ve_0,y+\ve_2,\psi+g(t),k),
\end{gather}
where $\ve_0$, $\ve_1$ and $\ve_2$ run through the set of real numbers,
and the parameter~$g$ runs through the set of real-valued smooth time-dependent functions. 
The corresponding Lie algebra is spanned by the vector fields 
$\p_t$, $\p_x$, $\p_y$, $\ZZ(g)$ trivially prolonged to~$k$.
Two kinds of transformations are missed in comparison with~$\bar G$,  
\begin{itemize}\itemsep=0.5ex
\item
the scale transformations $(\tilde t,\tilde x,\tilde y,\tilde\psi,\tilde k)=(e^{\ve_3}t,e^{-\ve_3}x,e^{-\ve_3}y,e^{-3\ve_3}\psi,e^{-4\ve_3}k)$ and 
\item
the generalized Galilean boosts $(\tilde t,\tilde x,\tilde y,\tilde\psi,\tilde k)=(t,x+f(t),y,\psi-f_t(t)y,k)$ with $f_t\ne0$. 
\end{itemize}
Note though that the  invariance under scale transformations is implicitly restored in~\cite{mars10Ay} by extending the scale transformation to also act on the hyperdiffusion coefficient~$A$. This will be discussed in detail below, were we present an alternative form for hyperdiffusion that will allow us to keep $A$ dimensionless. In doing this, we follow~\cite{bihl11Fy} where it was shown that classical biharmonic diffusion (and, more generally, any kind of \emph{linear} hyperdiffusion) breaks the scale invariance of the vorticity equation on the beta-plane. Galilean invariance is violated due to the presence of the first term on the right-hand side in the eddy-kinetic energy equation in system~\eqref{eq:MarshallSystem}. Note that this term is central for the dynamics of the barotropic ocean model, since it models the energy conversion between the eddies and the mean field~\cite{mars10Ay}.

As a way to recover the invariance of system~\eqref{eq:MarshallSystem}, we could use the moving frame method to invariantize these two equations using the moving frame associated with the maximal Lie invariance group of the vorticity equation. This would lead to a system that is by construction invariant under the same symmetry group. The downside of this approach is that the resulting invariant system of equations would depend on the moving frame to invariantize the original system, and most of moving frames lead to invariantizations that quite differ in form and properties from the original system. As there are infinitely many moving frames, consequently also infinitely many invariantized systems of equations could be constructed. In practice, of course all of these systems can be related by considering suitable combinations of each system with differential invariants from the algebra of differential invariants. As this can be a potentially tedious endeavor, we proceed in a more direct fashion here.

\looseness=1
In particular, we argue that Galilean invariance is not of relevance for the parameterized ocean model and only scale invariance needs to be restored. The argument for this is that in most numerical experiments system~\eqref{eq:MarshallSystem} will be used, and in particular in all the experiments carried out in Section~\ref{sec:numericalResultsBarotropicOcean}, the domain of integration will remain at rest and fixed Dirichlet boundary conditions will be used. The presence of fixed boundaries breaks both the scale invariance and the Galilean invariance of the original vorticity equation (ultimately, such boundaries break all symmetries except for time translations and gauging of the stream function), see e.g.~\cite{bihl11By}. We then follow the interpretation of symmetries as \textit{equivalence transformations}, which was originally proposed in~\cite{bihl12By} and is in line with the above concept of generalized invariant parameterization.

There it was argued that when transforming a given reference frame to another reference frame, also the relevant initial-boundary value problems of both frames should be mapped to each other. In other words, an appropriate symmetry transformation should map initial-boundary value problems from a given class of problems (e.g.\ Dirichlet boundary value problems varying the size of the spatial domain and the initial time, as well as the boundary values and initial conditions), to each other. The symmetries of the original system of differential equations then act as equivalence transformations of a particular class of initial-boundary value problems. Within this interpretation of the action of symmetries as equivalence transformations, the preservation of scale invariance is crucial since system~\eqref{eq:MarshallSystem} can be implemented on vastly different spatial and temporal scales, ranging from the standard, non-dimensional $2\pi\times 2\pi$ spatial domain~\cite{wang94a} to the characteristic ocean scale spatial domain of $4000\ \rm{km}\times4000\ \rm{km}$~\cite{mars10Ay}. As such, the inclusion of scale invariance in any parameterization for this model is crucial to ensure the various sized domains are appropriately mapped to each other. 
The scale factor of standard hyperdiffusion differs from those of the other terms in the first equation of system~\eqref{eq:MarshallSystem} and hence the parameter~$A$ varies within the class of initial-boundary value problems usually considered. On the other hand, since the typically imposed Dirichlet boundary conditions (see Section~\ref{sec:numericalResultsBarotropicOcean}), are not natural for constantly moving reference frames, the inclusion of Galilean invariance in a parameterized ocean model is not natural, and hence will not be pursued below.

For later use, we also introduce the essential subgroup $\bar G_{\rm ess}$  that consists of the transformations
\[
 (\tilde t,\tilde x,\tilde y,\tilde\psi,\tilde k)=(e^{\ve_3}(t+\ve_1),e^{-\ve_3}(x+\ve_0),e^{-\ve_3}(y+\ve_2), e^{-3\ve_3}(\psi+g(t)), e^{-4\ve_3}k),
\]
i.e., which includes all elementary transformations from~$\bar G$, except the generalized Galilean boosts with non-constant values of $f$.

To restore the scale invariance of system~\eqref{eq:MarshallSystem}, it is possible to use the results derived in~\cite{bihl11Fy} and to extend them to the present case. This is readily accomplished by noting that the terms on the left-hand side of the vorticity equation in~\eqref{eq:MarshallSystem} scale as $e^{-2\ve_3}$; consequently, also the terms on the right-hand side should scale as~$ e^{-2\ve_3}$. Let us thus investigate the scaling properties of the terms $\nn\cdot(\kappa\nn\eta)$ and $A\nn^4\eta$.

For the first term $\nn\cdot(\kappa\nn\eta)$, if we assume the relation $\kappa = \alpha L_{\rm eddy} (2k)^{1/2}$ and suppose that the eddy mixing length scale $L_{\rm eddy}$ scales the same as $x$ and $y$, i.e.\ $L_{\rm eddy}\sim e^{-\ve_3}$, then this term indeed scales as~$e^{-2\ve_3}$. As for the second term $A\nn^4\eta$, it scales as $e^{3\ve_3}$, i.e.\ the constant $A$ cannot be dimensionless (and in practice it is not, but rather is chosen to depend on the grid spacing, see e.g.~\cite{mars10Ay} and Section~\ref{sec:numericalResultsBarotropicOcean} below). This is the problem reported in~\cite{bihl11Fy}, i.e.\ linear hyperdiffusion cannot preserve the scale invariance of the vorticity equation.

A possible correction of this violation of scale invariance is achieved by using $\tilde A|\psi_x|^{5/2}\nn^4\eta$ instead of linear hyperdiffusion, where $\tilde A$ now \emph{is} dimensionless. An invariant form of the parameterized vorticity equation is therefore
\begin{equation}\label{eq:InvariantBiharmonicDiffusion}
 \eta_t + \psi_x\eta_y - \psi_y\eta_x = \nn\cdot(\kappa\nn\eta)-\tilde A|\psi_x|^{5/2}\nn^4\eta.
\end{equation}
If $\kappa$ was put to zero (realized by setting $k=0$, leading back to the standard first-order closure), the above form of the vorticity equation is identical with the one used in~\cite{bihl11Fy} to carry out freely decaying turbulence simulations on the beta-plane. It was advocated in~\cite{bihl11Fy} that the coefficient $|\psi_x|^{5/2}$ in the hyperdiffusion is necessary in order to guarantee that the constant $\tilde A$ is really dimensionless.

Here we have more choices for an appropriate scaling coefficient in the hyperdiffusion term, because we carry along an explicit equation for the turbulent kinetic energy. Thus, an alternative invariant form of a closed vorticity equation is
\begin{equation}\label{eq:InvariantTurbulentEnergyDiffusion}
 \eta_t + \psi_x\eta_y - \psi_y\eta_x = \nn\cdot(\kappa\nn\eta)-\tilde Ak^{5/4}\nn^4\eta.
\end{equation}
It can be verified by direct computation that the equation~\eqref{eq:InvariantTurbulentEnergyDiffusion} is invariant under all the transformations from the group~$\bar G$. 
In particular, $\tilde A$ is a true, dimensionless constant that needs no scaling to guarantee the invariance of the equation. Moreover, in~\cite{mars10Ay} the parameter $A$ was chosen to be proportional to $\Delta x^4$, i.e.\ as the mesh is refined, the strength of the hyperdiffusion is reduced accordingly. Setting $A=\tilde Ak^{5/4}$ has the same effect as the strength of the turbulent kinetic energy is reduced the finer the resolution of the model is. In other words, (21) uses a self-refining parameterization scheme.

\begin{remark}
An alternative to the parameterization proposed in~\eqref{eq:InvariantTurbulentEnergyDiffusion} would be
\[
\eta_t + \psi_x\eta_y - \psi_y\eta_x = \nn\cdot(\kappa\nn\eta)-\nn^2(\tilde Ak^{5/4}\nn^2\eta).
\]
This form is still invariant under all the transformations from the group~$\bar G$. From the physical point of view one reason for choosing this particular form for a hyperdiffusion-like term would be that diffusion is now in \textit{conserved form}. That is, as it stands, the above equation is still a conservation law for the vorticity, i.e.\ the parameterized system preserves the total circulation. See~\cite{bihl11Fy,bihl12Dy,popo17a} for a more in-depth discussion of such \emph{invariant and/or conservative parameterization schemes}. The downside of this parameterization is that the nonlinear hyperdiffusion term becomes even more complex than the one given in~\eqref{eq:InvariantTurbulentEnergyDiffusion}. For this reason, we will continue with the form~\eqref{eq:InvariantTurbulentEnergyDiffusion} in the sequel. 
Various alternatives to~\eqref{eq:InvariantTurbulentEnergyDiffusion} will be tested in further papers. 
\end{remark}

\looseness=1
Let us now turn to the closure given in~\cite{mars10Ay} for the kinetic energy equation~\eqref{eq:turbulentKineticEnergyEquation}, which is the second equation of system~\eqref{eq:MarshallSystem}. It is straightforward to check that the terms on the left-hand side of this equation scale as $k^{-5}$. To obtain a proper scaling for the terms of the right-hand side, the parameters~$\kappa$ and~$\nu$ have to scale like $e^{-3\ve_3}$, whereas the constant $r$ has to scale like $e^{-\ve_3}$. As we focus on the case of freely decaying turbulence hereafter, we in the following will omit the last term in the second equation of system~\eqref{eq:MarshallSystem}, i.e.\ we will be setting $r=0$. 
It is the second equation that significantly reduces the Lie symmetry group of system~\eqref{eq:MarshallSystem} to~$\bar G_{\rm min}$. 
In particular, this equation is not invariant with respect to the scale transformations from~$\bar G$. 

The remaining problem is thus to find functional expressions for $\kappa$ and $\nu$ such that the energy equation in system~\eqref{eq:MarshallSystem} remains invariant under the action of the scale transformation at least in the sense of equivalence transformations. Here, choosing $\kappa$ as defined in~\eqref{eq:ParameterKappa} would yield a proper scaling provided that we scale $L_{\rm eddy}\sim e^{-\ve_3}$ as indicated above. For the sake of simplicity, we choose for $\nu$ a similar form as was chosen for~$\kappa$ and set
\begin{equation}\label{eq:ParameterKappaE}
 \nu = 2\tilde\alpha L_{\rm eddy}(2k)^{1/2},
\end{equation}
with $\tilde \alpha$ being another dimensionless constant.

The invariant one-and-a-half closure model for the averaged barotropic vorticity equation without linear energy decay therefore reads
\begin{align}\label{eq:MarshallSystemInvariant}
\begin{split}
 &\eta_t + \psi_x\eta_y - \psi_y\eta_x = \nn\cdot(\kappa\nn\eta)-\tilde Ak^{5/4}\nn^4\eta,\\
 &k_t + \psi_xk_y - \psi_yk_x = -\kappa \nn\psi\cdot\nn\eta + \nn\cdot(\nu\nn k),
\end{split}
\end{align}
where $\kappa$ and $\nu$ are given in~\eqref{eq:ParameterKappa} and~\eqref{eq:ParameterKappaE}, respectively. 
The family of systems of the form~\eqref{eq:MarshallSystemInvariant}, 
where $\alpha$, $\tilde A$ and $\tilde\alpha$ are assumed to be fixed constants, 
and $L_{\rm eddy}$ is a varying constant parameter, is a class of systems of differential equations 
with~$L_{\rm eddy}$ playing the role of the arbitrary element. 
Each system of the class is invariant with respect to the group~$\bar G_{\rm min}$, 
which is the proper subgroup of~$\bar G_{\rm ess}$ singled out by the constraint $\ve_3=0$. 
Moreover, the equivalence group of the class contains the prolongation of~$\bar G_{\rm ess}$ to~$L_{\rm eddy}$ 
via $\tilde L_{\rm eddy}=e^{-\ve_3}L_{\rm eddy}$.
As a result, the above class gives the first example for an invariantly closed system employing a closure of order higher than one, 
which preserves, in the generalized sense, the subgroup of Lie symmetries of the barotropic vorticity equation on the beta-plane 
that is natural for a class of Dirichlet boundary value problem. 

\begin{remark}
An alternative form for the eddy transfer coefficient $\kappa$ found in the literature (see e.g.~\cite{eden08Ay,mars10Ay}) is
 $
  \kappa= 2\gamma \mathcal T_{\rm eddy} k,
 $
where $\gamma$ is a dimensionless constant and $\mathcal T_{\rm eddy}$ is a the eddy turnover timescale. It is readily verified that this choice for the parameter $\kappa$ would be scale invariant for the vorticity equation, i.e.\ $\gamma$ would really be a dimensionless constant.
\end{remark}

Before we move on with the presentation of numerical results for the model~\eqref{eq:MarshallSystemInvariant} we should like to discuss more specifically a novel interpretation of parameterization constants in the framework of symmetry-preserving parameterization schemes, an example for which was given above. A key observation is that in the definition of $\kappa$ as employed in~\eqref{eq:ParameterKappa} we identify the parameter $L_{\rm eddy}$ as \emph{non-dimensionless}. This is fully compatible with the physical interpretation of this parameter, which is the eddy mixing length. As we re-scale the barotropic vorticity equation using the scale transformation $\Gamma_\ve\colon (\tilde t,\tilde x,\tilde y,\tilde \psi,\tilde k) = (e^{\ve}t,e^{-\ve}x,e^{-\ve}y, e^{-3\ve}\psi, e^{-4\ve}k)$ we scale the time-space domain of the problem as well. It is therefore natural to extend the action of the scale operation to the eddy mixing length (and eddy turnover time) as well. In mathematical terms, we let the symmetry transformation $\Gamma_\ve$ act as an equivalence transformation on $L_{\rm eddy}$. In other words, we regard system~\eqref{eq:MarshallSystem} with $r=0$, the definition of~$\kappa$ given in~\eqref{eq:ParameterKappa} and $\nu$ given~\eqref{eq:ParameterKappaE}, as a class of systems of differential equations, where~$L_{\rm eddy}$ is a constant arbitrary element and~$A$ is a fixed proper relative differential invariant of the one-parameter scale group ($\Gamma_\ve$, $\ve\in\mathbb{R}$). The scale transformation $\Gamma_\ve$ acts as equivalence transformation in this class, meaning it maps one system from the class (with a given fixed value of $L_{\rm eddy}$) to another system from the same class (with another value of $L_{\rm eddy}$). For an extensive review of equivalence transformations in the framework of group classification, see e.g.\cite{bihl11Dy,ovsi82Ay,popo06Ay,popo10Ay}.

The situation is comparable to the construction of invariant discretization schemes as discussed in~\cite{bihl12By}. Here it is convenient to allow symmetries to act as equivalence transformations on the given boundaries of the domain of integration, because requiring symmetries to actually leave invariant these boundaries is both overly restrictive (meaning the corresponding symmetry group would be often trivial) and not natural from the physical point of view.

The analogous interpretation of symmetry transformations in the framework of invariant parameterization schemes is therefore both physically justified and convenient from the practical point of view. It can be much easier to accommodate a re-scaling of the physical parameters of the problem rather than having to introduce a combination of differential functions to create an invariant expression.

Nevertheless, care should be taken with this new interpretation of symmetries acting as equivalence transformations in the parameterization problem. For example, it would not be possible to avoid the multiplier $|\psi_x|^{5/2}$ in the invariant biharmonic diffusion derived in~\cite{bihl11Fy} and used in~\eqref{eq:InvariantBiharmonicDiffusion} by a mere re-scaling of the parameter $\tilde A$. The reason for this is that there is no obvious physical justification which would require the re-scaling of $\tilde A$ once the transformation $\Gamma$ is applied to the system~\eqref{eq:MarshallSystem}. 
The same claim is relevant for the multiplier $k^{5/4}$ in the invariant biharmonic diffusion in the equation~\eqref{eq:InvariantTurbulentEnergyDiffusion}.

The above discussion extends naturally to symmetries other than scale transformation. In general, the possibility of relaxing the rather strict requirement of symmetry preservation in the parameterization problem to the possibility of letting symmetries act as equivalence transformation on the physical parameters arising in most closure schemes gives therefore a practicable new way for the construction of invariant parameterization schemes.

\begin{remark}
Note that there is a quite extensive history for the use of nonlinear (hyper)diffusion for large eddy simulations in both three-dimensional and two-dimensional (geophysical) fluid mechanics, see e.g.\ the seminal paper \cite{smag63a} and~\cite{leit96a} as well as the review paper~\cite{fox08a}. While the aforementioned results have been obtained based on considerations of the turbulent energy (resp.\ enstrophy) cascades, it is interesting to note that nonlinear (hyper)diffusion coefficients naturally arise within the paradigm of preserving symmetries in the filtered/averaged equations of fluid mechanics.
\end{remark}

\section{Numerical results}\label{sec:numericalResultsBarotropicOcean}

In this section we test the ability of system~\eqref{eq:MarshallSystemInvariant} to simulate the dynamics of a barotropic ocean. In particular, we focus on the ability of system~\eqref{eq:MarshallSystemInvariant} to simulate the emergence of the so-called \textit{Fofonoff vortices}~\cite{fofo54a}.

We have chosen this test problem for several reasons. Firstly, the Fofonoff solution arises as an unforced and undamped solution of the vorticity equation. It is an analytical solution of the vorticity equation with no normal flow boundary conditions~\cite{fofo54a}. The Fofonoff solution is characterized by a linear relationship between the absolute vorticity and the stream function, i.e., $\eta=\mu\psi+\lambda$ for some constants $\mu$ and $\lambda$. It has also been shown to be the equilibrium solution within the statistical mechanics consideration of two-dimensional geostrophic fluid mechanics, see e.g. the review article~\cite{bouc12a}. Moreover, the Fofonoff solution is also the maximum entropy solution under the constraint of minimum enstrophy~\cite{bret76a}. These various ways of deriving/interpreting the Fofonoff solution illustrate its supreme importance in geophysical fluid dynamics.

It was discussed in~\cite{wang94a} that the emergence of the Fofonoff solution in the long term numerical simulations of the barotropic vorticity equation largely depends on the proper choice of boundary conditions, although the form of the diffusion invoked also plays a role. In~\cite{mars10Ay} it was shown that the Fofonoff solution also arises in the parameterized model~\eqref{eq:MarshallSystem}. This is important since~\eqref{eq:MarshallSystem} already has two source terms in the vorticity equation, the parameterized eddy vorticity flux and the hyperdiffusion term. Here we will test if the invariant (nonlinear) hyperdiffusion diffusion introduced to restore scale invariance allows for the emergence of Fofonoff vortices. Comparison against the model using standard, non-invariant hyperdiffusion will be particularly instructive.

\subsection{Discretization details}

In this section we present a few numerical tests using the invariant parameterization scheme proposed above. For this aim, we discretize system~\eqref{eq:MarshallSystemInvariant} using a finite difference model similar to the one used in~\cite{bihl11Fy} for the study of invariant hyperdiffusion models.

Specifically, the nonlinear advection term in the vorticity equation is discretized using the Arakawa Jacobian operator~\cite{arak66Ay} with boundary treatment as proposed in~\cite{salm89a}. All the other spatial derivatives are discretized using centered finite differences in the interior of the domain, and second-order one-sided finite differences at the boundaries. Time-stepping is done using a second-order explicit Runge--Kutta method (the explicit trapezoidal rule), at a fixed time step of $\Delta t=2\cdot 10^{-3}$. The domain size is $L_x=L_y=2\pi$, and all quantities are non-dimensional. For the integration a total of $128\times128$ grid points are used. The linear change in the Coriolis parameter is set to $\beta=5$. Overall, the numerical setup is similar to the one used in~\cite{wang94a}.

The parameterization constants of the invariant and non-invariant models, are $\alpha=0.01$, $L_{\rm eddy}\in[2\pi/100,2\pi/20]$, $\nu\in[10^{-4},10^{-3}]$, $\tilde \alpha=\nu/L_{\rm eddy}$ and $A\in[10^{-7},10^{-5}]$, respectively. Note that for most parameterization constants we list here a quite wide parameter range. We then choose various parameter combinations from these ranges to assess the performance of the invariant and non-invariant models in a statistical way.

\begin{figure}[!ht]
\centering
\includegraphics[scale=0.5]{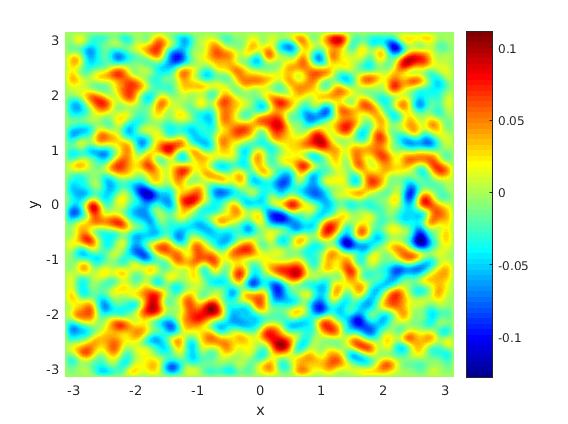}
\caption{The initial stream function for all experiments.}
\label{fig:initialStreamFunction}
\end{figure}

The initial stream function is generated by assigning a Gaussian distributed random number with zero mean and standard deviation $\sigma=0.25$ at each grid point, and subsequently replacing the value of the stream function at the grid point by its mean over the neighboring grid points. This cycle of replacing the current value of the stream function at each grid point by its mean over its neighbors is repeated 10 times, to give the initial stream function depicted in Fig.~\ref{fig:initialStreamFunction}.

\looseness=1
The turbulent kinetic energy is initialized to be constant $k=k_0$ throughout the entire domain. To facilitate comparison between the invariant and non-invariant runs, noting that $A=\tilde A k^{5/4}$, we set $\tilde A=A/\sqrt[4]{k_0^5}$. This guarantees that the hyperdiffusion strength of the invariant and non-invariant simulations are roughly of the same size, at least at the initial time. As the overall turbulent kinetic energy does not vary greatly over the integration period, this means that the diffusion strength is also similar over the entire integration, with the main difference being the nonlinearity of the hyperdiffusion term in the invariant case. We let the initial turbulent kinetic energy range over $k_0\in[0.10,0.25]$, which yields sufficiently strong initial eddy kinetic energies.

The boundary conditions for the vorticity equation are those of no normal follow, $\psi=0$, for the stream function and the superslip boundary conditions for the hyperdiffusion term, i.e., $\zeta_{\mathbf n}=0$ and $\zeta_{\mathbf n\mathbf n\mathbf n}=0$, where the subscript~$\mathbf n$ denotes the derivative in the normal direction to the boundary, $\zeta_{\mathbf n}=\mathbf n\cdot\nabla\zeta$; see~\cite{cumm92a,wang94a} for further discussions. For the eddy kinetic energy equation we use no flux boundary conditions, similar to what was used in~\cite{mars10Ay} and in~\cite{grea00a}.

\subsection{Results}

We integrate the discretized invariant system~\eqref{eq:MarshallSystemInvariant} and the discretized non-invariant system~\eqref{eq:MarshallSystem} until $t=500$. Following~\cite{wang94a}, as an indicator for approaching the Fofonoff solution we introduce the \textit{anti-correlation function}
\begin{equation}\label{eq:CorrelationFunction}
C=-\int_\Omega\zeta\beta y\,\mathrm{d}A,
\end{equation}
which measures the anti-correlation between the relative vorticity and the latitude. Here, $\Omega$ is the domain of integration in the plane and $\mathrm{d}A$ is the associated area element. A random vorticity field gives correlations close to zero. Since the Fofonoff solution is characterized by two vortices of opposite sign at the northern and southern parts of the oceanic basin, the anti-correlation function~$C$ will be positive. The increase of $C$ over the integration period starting with uncorrelated vorticity and stream function fields, is thus an indicator for the emergence of the Fofonoff solution.

A total of 72 runs are carried out, 36 runs at fixed $\nu=10^{-3}$ and 36 runs at fixed $\nu=10^{-4}$, with $L_{\rm eddy}$, $A$ and $k_0$ ranging over the intervals given above. We then measure the anti-correlation function at the end of the integration period.

Using the invariant parameterization model~\eqref{eq:MarshallSystemInvariant}, we find a mean final anti-correlation function of $\bar C_{\rm inv}=469$ (with standard deviation $\sigma_{\rm inv}=51$), whereas using the standard, non-invariant parameterization model~\eqref{eq:MarshallSystem} we find a mean final anti-correlation function of $\bar C_{\rm std}=445$ (with standard deviation $\sigma_{\rm std}=66$). In addition to slightly higher anti-correlation values and slightly less scattering results when varying the parameterization parameters, overall it seems that the use of the invariant parameterization model leads to more stable numerical integrations of the freely decaying turbulence test. The integration became unstable in 4 of the non-invariant simulations but in none of the invariant simulations.

\begin{figure}[!ht]
\centering
\begin{subfigure}{.5\textwidth}
  \centering
  \includegraphics[width=1\linewidth]{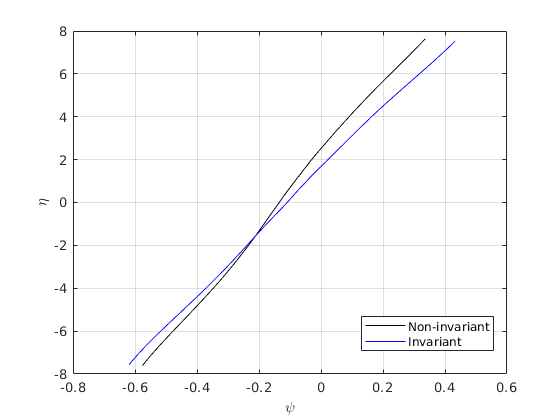}
\end{subfigure}%
\begin{subfigure}{.5\textwidth}
  \centering
  \includegraphics[width=1\linewidth]{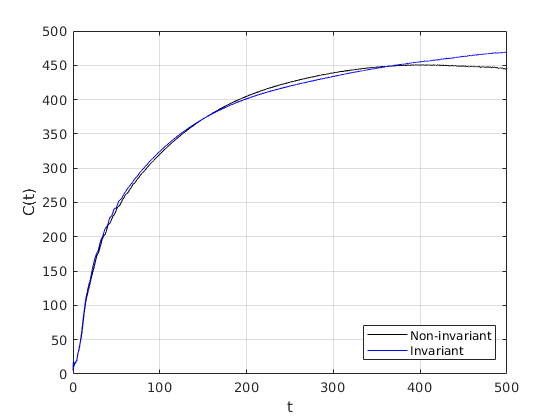}
\end{subfigure}
\begin{subfigure}{.5\textwidth}
  \centering
  \includegraphics[width=1\linewidth]{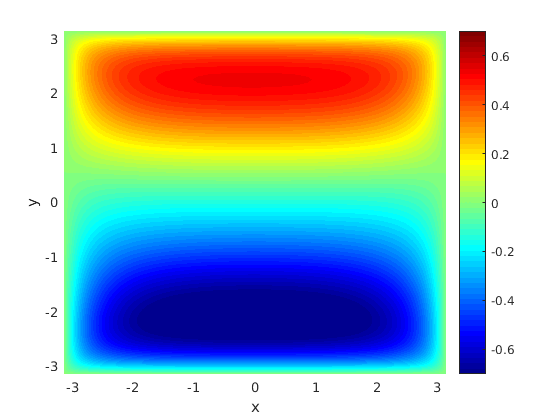}
\end{subfigure}%
\begin{subfigure}{.5\textwidth}
  \centering
  \includegraphics[width=1\linewidth]{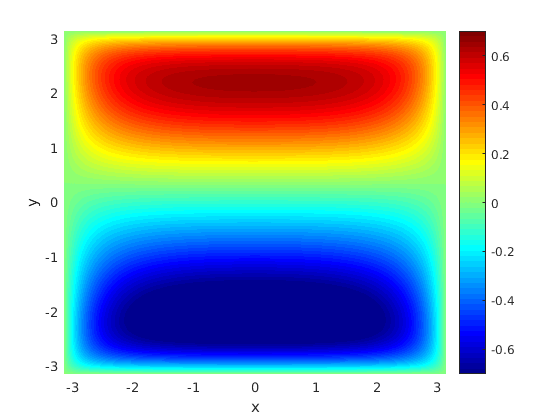}
\end{subfigure}
\caption{\textbf{Upper panel:} Zonally-averaged $\psi$--$\eta$ scatter plot (left) at $t=500$ and time series of the anti-correlation function (right) that are averaged over 72 runs varying the parameterization constants over the intervals $L_{\rm eddy}\in[2\pi/20,2\pi/100]$, $\nu\in[10^{-3},10^{-4}$] and $A\in[10^{-5},10^{-7}]$, with $\alpha=0.01$, $\tilde \alpha=\nu/L_{\rm eddy}$, and varying the initial turbulent kinetic energy over $k_0=[0.1,0.25]$. \textbf{Lower panel:} Final stream function at $t=500$ averaged over all 72 using the non-invariant model (left) and the invariant model (right).}
\label{fig:FofonoffSolutionAverage}
\end{figure}

The scatter plot of the absolute vorticity zonally averaged over all 72 runs against the stream function zonally averaged over all 72 runs is depicted in Fig.~\ref{fig:FofonoffSolutionAverage} (upper panel, left). The averaged time series of the anti-correlation function over all 72 runs is shown in Fig.~\ref{fig:FofonoffSolutionAverage} (upper panel, right). The former plot shows that the invariant model accurately represents the Fofonoff solution over the entire tested parameter range, whereas the non-invariant model is not able to capture the perfect linear relation between absolute vorticity and stream function. The latter plot shows that while the anti-correlation function has peaked for the non-invariant model average at about $t=350$ and started to decay afterwards, this function continues to rise for the invariant model over the entire integration period. This result is also consistent with the final stream function fields averaged over all runs, which are depicted in Figure~\ref{fig:FofonoffSolutionAverage} (lower panel), showing that the non-invariant model (lower panel, left) produces slightly weaker, and more asymmetric vortices than the invariant model (lower panel, right) does.

\begin{figure}[!ht]
\centering
\begin{subfigure}{.5\textwidth}
  \centering
  \includegraphics[width=1\linewidth]{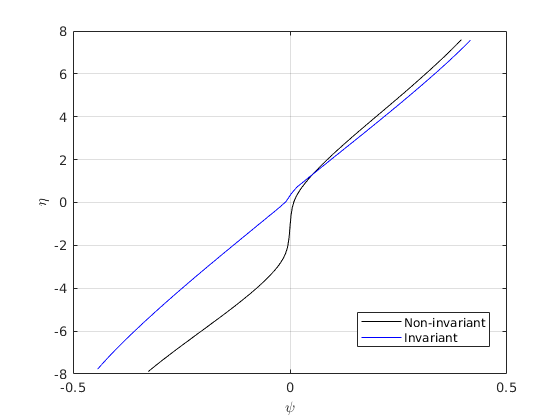}
\end{subfigure}%
\begin{subfigure}{.5\textwidth}
  \centering
  \includegraphics[width=1\linewidth]{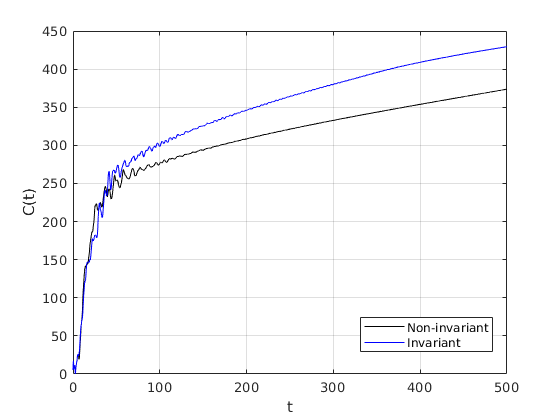}
\end{subfigure}
\vspace{1ex}
\begin{subfigure}{.5\textwidth}
  \centering
  \includegraphics[width=1\linewidth]{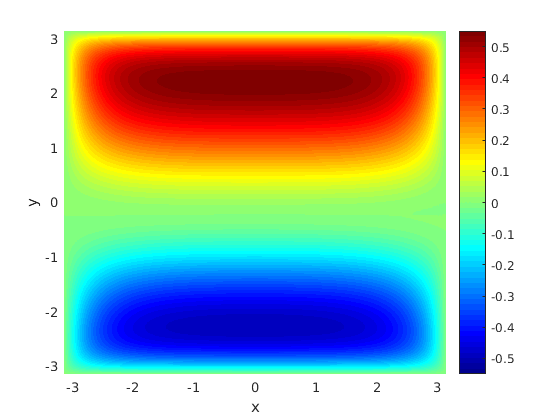}
\end{subfigure}%
\begin{subfigure}{.5\textwidth}
  \centering
  \includegraphics[width=1\linewidth]{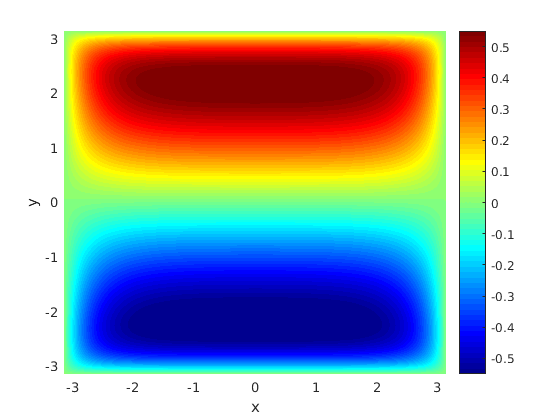}
\end{subfigure}
\caption{\textbf{Upper panel:} Zonally-averaged $\psi$--$\eta$ scatter plot at $t=500$ (left) and time series of the anti-correlation function (right). \textbf{Lower panel:} Final stream function at $t=500$ using standard hyperdiffusion (left) and invariant hyperdiffusion (right). The parameterization constants are $L_{\rm eddy}=2\pi/100$, $A=10^{-7}$, $\nu=10^{-3}$, starting with uniform initial turbulent kinetic energy set to $k_0=0.15$.}
\label{fig:FofonoffSolutionCharacteristic}
\end{figure}

To show some individual results for both models, in Fig.~\ref{fig:FofonoffSolutionCharacteristic} we present the scatter plot of the zonally averaged absolute vorticity $\eta$ over the zonally-averaged stream function $\psi$, as well as the time series of the anti-correlation function (upper panel), and the final stream functions for both the non-invariant and the invariant model (lower panel). The parameterization constants were set to $L_{\rm eddy}=2\pi/100$, $A=10^{-7}$, $\nu=10^{-3}$, starting with the uniform initial turbulent kinetic energy set to $k_0=0.15$.

Both the non-invariant and invariant model lead to the emergence of a Fofonoff-like solution, with the vortices being slightly stronger in the invariant model. Note that the non-invariant model homogenizes the stream function field between the two vortices, whereas no such homogenization takes place for the invariant model. This homogenization of the stream function field shows up as a kink in the scatter plot for the non-invariant model, whereas the invariant model reproduces the required linear relation between absolute vorticity and stream function perfectly. This result is further corroborated by the higher values of the anti-correlation function for the invariant model.

\begin{figure}[!ht]
\centering
\begin{subfigure}{.5\textwidth}
  \centering
  \includegraphics[width=1\linewidth]{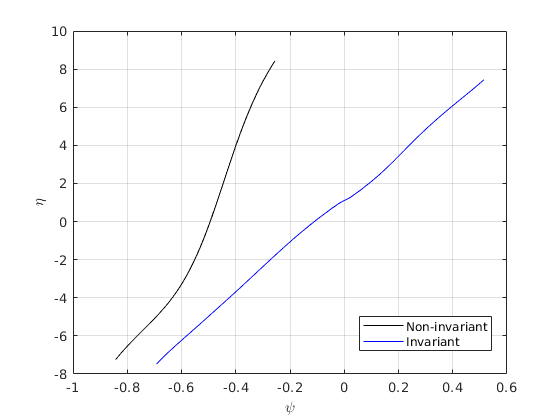}
\end{subfigure}%
\begin{subfigure}{.5\textwidth}
  \centering
  \includegraphics[width=1\linewidth]{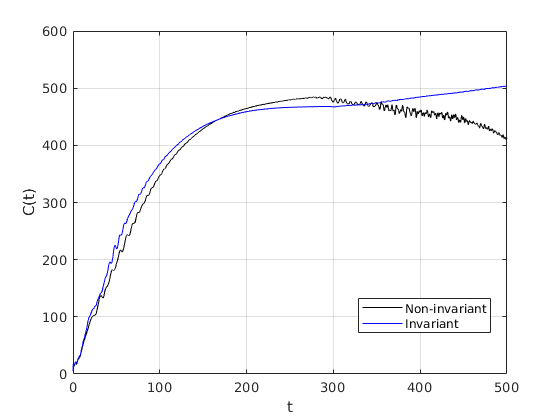}
\end{subfigure}
\vspace{1ex}
\begin{subfigure}{.5\textwidth}
  \centering
  \includegraphics[width=1\linewidth]{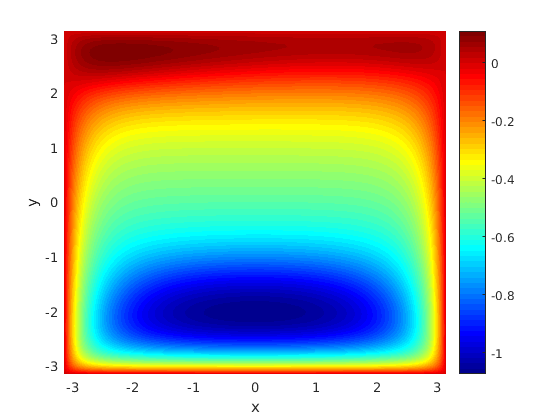}
\end{subfigure}%
\begin{subfigure}{.5\textwidth}
  \centering
  \includegraphics[width=1\linewidth]{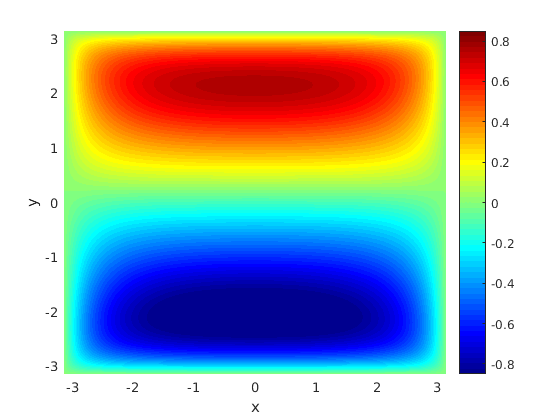}
\end{subfigure}
\caption{\textbf{Upper panel:} Zonally-averaged $\psi$--$\eta$ scatter plot at $t=500$ (left) and time series of the anti-correlation function (right). \textbf{Lower panel:} Final stream function at $t=500$ using standard hyperdiffusion (left) and invariant hyperdiffusion (right). The parameterization constants are $L_{\rm eddy}=2\pi/20$, $A=10^{-6}$, $\nu=10^{-4}$, starting with uniform initial turbulent kinetic energy set to $k_0=0.15$.}
\label{fig:FofonoffSolutionCharacteristic2}
\end{figure}

As another individual result, in Fig.~\ref{fig:FofonoffSolutionCharacteristic2} we depict the zonally-averaged $\psi$--$\eta$ scatter plots and time series of the anti-correlation function (upper panel), and the plots of the respective final stream functions (lower panel), for the parameterization constants $L_{\rm eddy}=2\pi/20$, $A=10^{-6}$, $\nu=10^{-4}$, starting with the uniform initial turbulent kinetic energy again set to $k_0=0.15$.

Again the invariant model produces the required linear relation between $\eta$ and $\psi$ almost perfectly, whereas in the non-invariant model the Fofonoff vortices start to decay. This is also visible in a drop of the anti-correlation function starting at $t=300$, which previously stabilized at values near $C=500$. In contrast, in the invariant model the anti-correlation function keeps increasing over the entire integration period.

Both the previous and these results are consistent with those presented in~\cite{duko99a}, where it was pointed out that Fofonoff vortices will be converted to homogenized gyres provided the diffusion is strong enough and the integration interval is long enough. It seems that the invariant parameterization model is capable of holding off the homogenization tendency longer in the present case and thus stabilizes the Fofonoff solution.

Overall, the results obtained show that the invariant model~\eqref{eq:MarshallSystemInvariant} is capable of robustly producing the Fofonoff-type solutions for a widely varying range of the parameterization constants and initial values of the eddy kinetic energy.

\section{Conclusion}\label{sec:conclusionBarotropicOcean}

\looseness=-1
This paper was devoted to the problem of invariant parameterization schemes for eddies in a barotropic ocean model. In particular, we showed that it is possible to construct invariant closure schemes not only of order one but also of higher order. For the purpose of demonstration, we started with the one-and-a-half order closure model proposed in~\cite{mars10Ay} and turned it into a model that preserves, in the generalized sense, an essential subgroup of the Lie symmetry group of the original incompressible Euler equations in stream-function form on the beta-plane, which is the barotropic vorticity equation.  
One of the main observations in the present paper was that any Lie symmetry of this equation is naturally prolonged to the kinetic energy. 
Therefore, the Lie symmetry group of the single barotropic vorticity equation is canonically isomorphic to the Lie symmetry group of the joint system of this equation and the kinetic energy equation, and the latter system was parameterized in~\cite{mars10Ay}. 
To take into account the turbulent kinetic energy in invariant parameterization schemes of order one-and-a-half, it was necessary to prolong the action of the Lie symmetry group of the vorticity equation to the turbulent kinetic energy. This was done by examining the action of this group on the first spatial derivatives of the stream function and splitting them into a mean and a perturbational part. This yielded that only the prolongations of scale transformations act non-identically on the dependent variable associated with the turbulent kinetic energy. Carrying out this prolongation procedure is another main new features of this paper. 

We also showed that not all Lie symmetries the barotropic vorticity equation should be taken into account in the course of the construction of invariant parameterization schemes. 
The symmetries that are essential for the construction constitute a proper subgroup of the maximal Lie symmetry group of the original vorticity equation. 
We singled out this subgroup via analysis of related initial-boundary value problems, and it is the prolongation of this subgroup that parameterized models employing a closure should preserve in certain sense. 

Another new observation reported in this paper was the interpretation of the invariant parameterization problem in the generalized sense. 
It can be natural for symmetries of models to be parameterized to act, after prolonging to subgrid-scale fields and to certain closure's parameters, as equivalence transformations within a set of selected closed averaged models. 
The reason for this is that most subgrid-scale closure schemes contain free constant parameters that can be associated to the physics of the process to be parameterized. When applying a symmetry transformation to closure model it may thus be appropriate to let the transformation act on these constants as well. In the field of group analysis of differential equations, such transformations are referred to as equivalence transformations. Having the possibility to extend symmetries of the original systems of unaveraged differential equations to equivalence transformations in the closed systems of averaged differential equations parallels the same observation for invariant discretization schemes recently established in~\cite{bihl12By}.

We have shown using numerical simulations that the proposed invariant closure model for barotropic eddies in the ocean is capable of producing the Fofonoff vortex solution for sufficiently long integration. In particular, we observe the organization of Fofonoff-type vortices out of a random, small-vortex dominated initialized stream function field. We have compared the invariant model, using invariant hyperdiffusion and a novel invariant energy-decay parameterization, against a non-invariant closure model and found the Fofonoff solution to be realized in a more robust way in the invariant model. Besides the results presented in~\cite{bihl11Fy}, where it was shown that the turbulent cascade in two-dimensional geostrophic turbulence on the beta-plane is obtained in a more robust way using invariant hyperdiffusion as well, this is the second example of an invariant parameterization outperforming a standard non-invariant one, showing the promise of the emerging field of geometric parameterization.

While we have presented the general theory for higher-order parameterization schemes in this paper as well, we reserve the problem of finding and testing closure models of order higher than one or one-and-a-half order for future investigations.

\section*{Acknowledgements}

The authors thank David Marshall for helpful remarks and discussions. 
This research was undertaken, in part, thanks to funding from the Canada Research Chairs program, the InnovateNL LeverageR{\&}D program and the NSERC Discovery program. 
AB is a recipient of an APART Fellowship of the Austrian Academy of Sciences.
The research of ROP was supported by the Austrian Science Fund (FWF), projects P25064 and P30233,
and by ``Project for fostering collaboration in science, research and education'' funded by the Moravian-Silesian Region, Czech Republic.

\footnotesize\setlength{\itemsep}{0ex}

\end{document}